# Ab-initio investigations for Structural, Mechanical, Optoelectronic, and Thermoelectric properties of Ba$_2$SbXO$_6$ (X=Nb, Ta) compounds

Hansraj[1,‡], K.C. Bhamu[2,3,*], Sung Gu Kang[3,*], A.K. Kushwaha[4,‡], D.P. Rai[5], Subrahmanyam Sappati[6], J. Sahariya[7,‡], Amit Soni[8,*]

[1]Department of Physics, Manipal University Jaipur, Jaipur-303007, Rajasthan, India
[2]Department of Physics, Gramin Mahila P.G. College, Sikar, Rajasthan, 332024, India
[3]School of Chemical Engineering, University of Ulsan, 93 Daehakro, Nam-Gu, Ulsan 44610, South Korea
[4]Department of Physics, K.N. Govt. P.G. College, Gyanpur, Bhadohi, 221304, India
[5]Department of Physics, Pachhunga University College, Aizawl 796001, India
[6]Raman Research Institute, C. V. Raman Avenue, Bengaluru 560080, India
[7]Department of Physics, National Institute of Technology, Uttarakhand, Srinagar (Garhwal)-246174, India
[8]Department of Electrical Engineering, Manipal University Jaipur, Jaipur-303007, Rajasthan, India
‡equal contribution; *Corresponding authors E-mail: kcbhamu85@gmail.com; sgkang@mail.ulsan.ac.kr; amitsoni_17@yahoo.co.in

**Abstract**

We report the structural, mechanical, electronic, optical, thermoelectric properties and spectroscopic limited maximum efficiency (SLME) of oxide double perovskite structure Ba$_2$SbNbO$_6$ and Ba$_2$SbTaO$_6$ compounds. All the investigations were performed through the first-principles density functional theory (DFT). The obtained values for the elastic constants reveal the mechanical stability of the studied compounds. The calculated data of bulk modulus (B), shear modulus (G), and Young's modulus (E) for Ba$_2$SbTaO$_6$ are found to be greater than those of Ba$_2$SbNbO$_6$. The ratio of Bulk to shear ratio (B/G) shows that Ba$_2$SbNbO$_6$ and Ba2SbTaO6 are ductile. The computed electronic band structure reveals the semiconducting nature of both compounds. We have also studied the electron relaxation time-dependent thermoelectric properties, such as Seebeck coefficient, thermal conductivity, electrical conductivity, thermoelectric power factor, and the figure of merit as a function of chemical potential at various temperatures for p-type and n-type charge carriers. The high absorption spectra and good figure of merit (ZT) reveal that both the studied compounds, $Ba_2SbXO_6$ (X = Nb, Ta) are promising materials for photovoltaic and thermoelectric applications. The calculated SLME of 26.8% reveals that $Ba_2SNbO_6$ is an appealing candidate for single-junction solar cells.

**1. Introduction**

The energy crisis due to the depletion of global energy resources has urged the researchers to deal with the alternate methods via utilization of energy from efficient solid-state materials. The utilization and harvesting of renewable sources like, solar energy, for sustainability from solid state device looks promising for this purpose. However, the discovery of efficient, economical and eco-friendly solid material is challenging task for the scientific community. In this regard, the perovskite compounds have been explored extensively well due to their potential applications in various technological fields such as, transistors, ferroelectric and piezoelectric materials, solar cell materials, light emitting devices, high temperature superconductors, magnetoresistances, photocatalytic and photovoltaics [1-8]. The prototype perovskites have oxide based structure along with large band gap. Therefore, these materials are not suitable for optoelectronic applications. In the last decade, many efforts have been made to develop organic-inorganic hybrid materials, i.e., the materials with the combination of an organic compound and lead halide perovskite, to enhance the efficiency of photo-electrochemical cells [9-11]. Yet, the main drawback of lead halide perovskite is the reduction of its efficiency because of moisture, temperature, etc. [12, 13]. The other disadvantage is related to the lead toxicity [14, 15].

Further, the migrations of halide vacancies also make these materials highly unstable [16, 17]. Therefore, the scientific community's focus has shifted to discover the materials that are eco-friendly, stable at ambient conditions, and hence can work more efficiently for a long time. To deal with this issue, some efforts were made to replace the Pb-atoms with Sn/Ge atoms but these atoms oxidize in the environment and hence the life time of devices with these compounds degraded [18-21].

To overcome these practical problems associated with these perovskite compounds, the researchers have explored double perovskite compounds that have a general formula of $A_2B'B''C_6$, where A site atoms may be the alkali metals, alkaline earth metals, or lanthanide series atoms, B-atoms may be the lanthanide series atoms or the *d*-block elements, and C-atoms are generally the oxygen or halogen atoms. The double perovskite compounds have shown their utilization in ferroelectric, ferromagnetic and many other applications [22-25]. In this context, McClure et al. [26] have synthesized the double perovskite structured $Cs_2AgBiX_6$ (X = Cl, Br and I) compounds and have determined the electronic and optical properties of the synthesized compounds. In their study, authors found these compounds are more efficient and stable than the other organic-inorganic materials and hence these are reported to be more suitable for the photo-electrochemical cells. Volonakis et al. [27] have studied the optoelectronic properties of double perovskite compounds $Cs_2B'B''X_6$ (B′ = Bi, Sb; B″ = Cu, Ag, Au; and X = Cl, Br, I) and found that these compounds have tunable band gaps within the visible range with low carrier effective mass. The double perovskite compounds $Cs_2InAgCl_6$, $Cs_2InAgBr_6$ and their mixed compound $Cs_2InAg(Cl_{1-x}Br_x)_6$ and $Cs_2AgIO_6$ have been synthesized very recently [28, 29]. A theoretical investigations the electronic properties of double-perovskites predicted that these compounds exhibit a direct semiconducting band gap and are potential renewable energy materials [28-30]. Zhou et al. [31]

have synthesized and studied the electronic and optical properties of Mn-doped $Cs_2NaBi_{1-x}In_xCl_6$. Authors found that the increase in doping concentration of Bi at the In site lead to the transition from indirect to direct bandgap nature.

All the above studies revealed that several research groups have made rigorous efforts to understand the core issue of the optoelectronic properties of the double perovskite $A_2BB'X_6$. The researchers are trying to replace A, B and B' with various atoms to find the variation in optoelectronic properties. In this regard, we have studied the structural, elastic, electronic, optical, thermoelectric, and thermodynamic properties of lead-free double perovskite compounds $Ba_2SbNbO_6$ and $Ba_2SbTaO_6$ based on the first-principles density functional theory (DFT).

## 2. Computational Details

The computations of structural, elastic, electronic, and optical properties of $Ba_2SbNbO_6$ and $Ba_2SbTaO_6$ compounds have been performed using Perdew-Burke-Ernzerhof (PBE) functional within the density functional theory (DFT) [32] using the Wien2K [33] code software based on the full-potential linearized augmented plane wave (FP-LAPW). Further, a more accurate modified Becke-Johnson (mBJ) exchange potential has also been used to obtain the more precise electronic properties of studied compounds [34]. We have also treated the interaction of the heavier atoms within spin-orbit coupling (SOC) as PBE+mBJ+SO for further improvement in the calculation of band structure. The elastic constants, optical properties, and thermoelectric properties have been calculated using PBE+mBJ potential. In FP-LAPW method, space is divided into two parts; one is the interstitial region, and the other is the non-overlapping muffin tin (MT) spheres . The MT radii were taken to be 2.45, 2.20, 1.87, 2.02 and 1.62 atomic units (a.u.) for Ba, Sb, Nb, Ta and O atoms, respectively. $K_{max}*R_{MT}$, cut-off value of plane wave, was set at 8, where $R_{MT}$ is the smallest muffin tin radii in the unit cell. The value for maximum Fourier expansion of charge density ($G_{max}$) was set as 12. The reciprocal space of the irreducible Brillouin zone was integrated using a total of 47 $k$-points with a mesh size of 10x10x10 grid. The self-consistent calculations were converged with respect to the total evergy convergence threshold of $10^{-5}$ Ry. To calculate the thermoelectric properties, we used BoltzTraP code [35]. It works under rigid band approximation (RBA) and constant relaxation time approximation (CRTA). Under RBA and CRTA approximations, we assume that the band structure remains unchanged with doping and only chemical potential changes and the relaxation time is independent of energy. For crystal structure visualization, we used VESTA software [36].

## 3. Results and Discussion

### 3.1 Structural properties

To start with the detailed investigations for the various properties of $Ba_2SbNbO_6$ and $Ba_2SbTaO_6$ compounds, we have optimized the lattice parameters for both the compounds using PBE functional. The optimized lattice parameters for $Ba_2SbNbO_6$ and $Ba_2SbTaO_6$ compounds are 8.581 Å and 8.589 Å, respectively. These double perovskite compounds crystallize in cubic structure with space group $Fm\bar{3}m$ where the Ba, Sb, Nb/Ta, O atoms occupies 8c (0.25, 0.25, 0.25), 4a (0, 0, 0), 4b (0.5, 0, 0) and 24e (x, 0, 0) sites respectively. The crystal structure of $Ba_2SbNbO_6$ is shown in Fig.1. The calculated lattice parameters of $Ba_2SbTaO_6$ are greater than the lattice parameters of $Ba_2SbNbO_6$ which may be due to the size effect of the atoms [37].

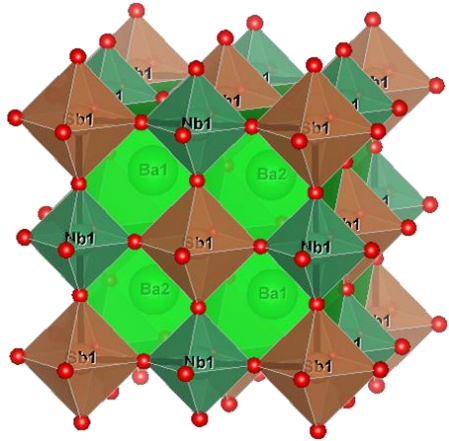

Fig. 1 Crystal structure for cubic $Ba_2SbNbO_6$ compound as a representative structure for both the studied compounds. Each atom in Fig. has labeled with the atom name except for the oxygen atom which is represented by a red color ball. For the $Ba_2SbTaO_6$ structure, the Ta atom will occupy the position of the Nb atom in the Figure.

## 3.2 Elastic properties

To engineer materials for industrial applications, elastic and mechanical properties play a critical role. These include second-order elastic constants ($C_{ij}$), Young's modulus, Poisson's ratio, bulk modulus, shear modulus, brittleness, ductility, anisotropy, propagation of elastic waves and other solid-state phenomena, etc. The three independent elastic constants $C_{11}$, $C_{12}$ and $C_{44}$, measures the mechanical stability of any system. These elastic constants are derived from some matrix of equations and should satisfy below born elastic stability criteria, for cubic system [38], $(C_{11}–C_{12})/2 > 0$, $(C_{11}+2C_{12}) > 0$, $C_{11} > 0$ and $C_{44} > 0$.

From Table-1, we see that our calculated values of $C_{11}$, $C_{12}$ and $C_{44}$ for $Ba_2SbNbO_6$ ($Ba_2SbTaO_6$) are 251.82 (267.72), 78.77 (78.14) and 61.90 (63.37) in GPa, respectively. These values satisfy the born elastic stability criteria, showing the mechanical stability of these compounds. Further, the value of Cauchy pressure, which is the difference between $C_{12}$ and $C_{44}$ [39], is positive for both the studied compounds. This show that our systems have metallic bonding.

Table-1: Elastic constants, Bulk modulus (B), Shear modulus (G), Young modulus (E), Pugh's ratio (B/G), Poisson's ratio ($v$) and factor of anisotropy (A) for $Ba_2SbXO_6$ (X = Nb, Ta)

| | Elastic Constants (GPa/unit) | | | | | |
|---|---|---|---|---|---|---|
| | $C_{11}$ | $C_{12}$ | $C_{44}$ | B | B' | E |
| **$Ba_2SbNbO_6$** | 251.82 | 78.77 | 61.90 | 136.45 | 4.54 | 188.89 |
| **$Ba_2SbTaO_6$** | 267.72 | 78.14 | 63.38 | 141.33 | 4.56 | 190.104 |
| | $G_R$ | $G_V$ | G | B/G | $v$ | A |
| **$Ba_2SbNbO_6$** | 69.85 | 71.75 | 70.80 | 1.93 | 0.28 | 0.72 |
| **$Ba_2SbTaO_6$** | 73.06 | 75.94 | 74.50 | 1.90 | 0.27 | 0.67 |

The isotropic bulk modulus (B) provides information regarding the resistance to the volume change through external pressure. To determine the Bulk modules, we used Voigt-Russel-Hills approximations [40-43] which use the $C_{11}$ and $C_{12}$ elastic constants. Mathematically it can be represented by the relation,

$$B = B_V = B_R = \frac{C_{11} + 2C_{12}}{3}. \quad (1)$$

The calculated bulk modulus and derivative of bulk modulus for $Ba_2SbNbO_6$ and $Ba_2SbTaO_6$ are listed in Table-1. The bulk modulus for $Ba_2SbNbO_6$ exhibits less volume change as compared to the $Ba_2SbTaO_6$. Hardness and resistance against reversible deformation can be understood in terms of shear modulus (G). The shear modulus can be calculated as follows [44],

$$G = \frac{G_V + G_R}{2}, \quad (2)$$

where, $G_V$ and $G_R$ are the upper and lower limit of shear modulus, respectively. $G_V$ and $G_R$ can be determined with the help of elastic constants [44]. Mathematically,

$$G_V = \frac{C_{11} - C_{12} + 3C_{44}}{5} \tag{3}$$

$$G_R = \frac{5C_{44}(C_{11} - C_{12})}{4C_{44} + 3(C_{11} - C_{12})} \tag{4}$$

The values for $G_V$, $G_R$, and G for the $Ba_2SbNbO_6$ and $Ba_2SbTaO_6$ compounds, obtained in present computations, are shown in Table-1. The higher value of shear modulus for $Ba_2SbTaO_6$ (74.50 GPa) in comparison to $Ba_2SbNbO_6$ (70.80 GPa) shows the more rigid nature of $Ba_2SbTaO_6$ than $Ba_2SbNbO_6$. The Young's modulus (E) for $Ba_2SbNbO_6$ and $Ba_2SbTaO_6$ compounds is calculated with the help of bulk modulus and shear modulus using the following relation[44],

$$E = \frac{9BG}{3B + G} \tag{5}$$

The obtained values for E are also included in Table-1. The E value for $Ba_2SbTaO_6$ is greater than that of $Ba_2SbNbO_6$ which leads to the higher resistance against uniaxial tension in comparison to $Ba_2SbNbO_6$. Further, for practical application of any compound in device fabrication, one needs to understand the ductile and brittle nature of compound. This can be explained by Pugh's ratio (B/G) and Poisson's ratio ($v$) [45]. The $v$ can be obtained using the following relation:

$$v = \frac{3B - E}{6B} \tag{6}$$

The physical properties like brittleness, ductility, direction, stiffness, etc., are decided from B/G ratio and $v$. If the values of B/G and $v$ are higher than 1.75 and 0.26, respectively [46, 47], the compound shows the ductile behavior and is suppose to be suitable for device fabrication. However, if the value of $v$ is lesser than 0.26, the compound shows brittle nature, and if $v$ is equal to 0.26, the compound is plastic. From Table-1 it is clear that the value of B/G and $v$ for $Ba_2SbNbO_6$ ($Ba_2SbTaO_6$) are 1.93 (1.90) and 0.28 (0.27), respectively, which show that both the studied compounds are ductile and hence are suitable for device fabrication. Further, 's's Zener's anisotropy factor "A" calculated from Eq. 7, is less than one and suggesting anisotropic nature of both the compounds [48, 49].

$$A = \frac{2C_{44}}{C_{11}-C_{12}} \qquad (7)$$

**3.3 Electronic properties**

In this section, we discuss the electronic properties of $Ba_2SbXO_6$ (X=Nb, Ta) compounds in terms of their electronic band structures, total and partial densities of states (DOS). Band structure and DOS for both the compounds have been computed using the PBE, PBE+mBJ and PBE+mBJ+SO functional. Energy gap obtained from various functional along with the available earlier data are collated in Table-2 and the corresponding energy band diagram for $Ba_2SbNbO_6$ and $Ba_2SbTaO_6$ compounds are presented in Fig. 2(A) and Fig. 3(A), respectively. Within PBE computation, energy band gap for $Ba_2SbNbO_6$ and $Ba_2SbTaO_6$ compounds are found to be 1.48 and 1.93 eV, respectively which increases to 1.80 eV and 2.45 eV while using mBJ+SO. Further from Table-2, it is found that the energy gap computed from the mBJ+SO calculations is in closer agreement with the available data in comparison to the other calculations. Further, from Fig. 2(A) and Fig. 3(A), it is observed that both the compounds have direct bandgap nature as the valence band maximum (VBM) and conduction band minimum (CBM) both are lying at same *k* point. We also present the partial and total DOS for $Ba_2SbNbO_6$ and $Ba_2SbTaO_6$ compounds in Fig. 2(B) and 3(B), respectively, using the mBJ+SO coupling for deeper exploration of energy bands.

Table-2: Energy band gap of $Ba_2SbXO_6$ (X = Nb, Ta) compounds obtained from different calculation in present study alongwith the available data.

| | Energy gap ($E_g$) in eV | | | |
|---|---|---|---|---|
| | **PBE** | **PBE+mBJ** | **PBE+ mBJ + SO** | **GW0[24]** |
| **$Ba_2SbNbO_6$** | 1.48 | 1.46 | 1.80 | 2.35 |
| **$Ba_2SbTaO_6$** | 1.93 | 1.75 | 2.45 | 2.45 |

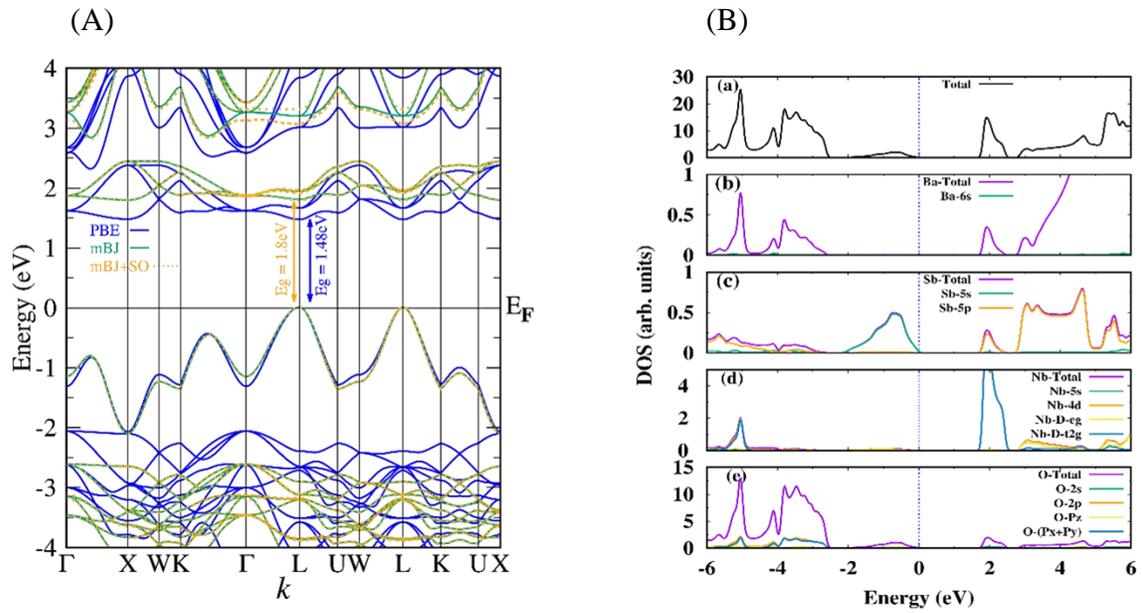

Fig. 2 (A) Energy bands for $Ba_2SbNbO_6$ computed using PBE, mBJ and mBJ +SO methods along the high symmetry directions of first Brillouin zone. (B) Total and partial density of states of $Ba_2SbNbO_6$ obtained from mBJ+SO potential.

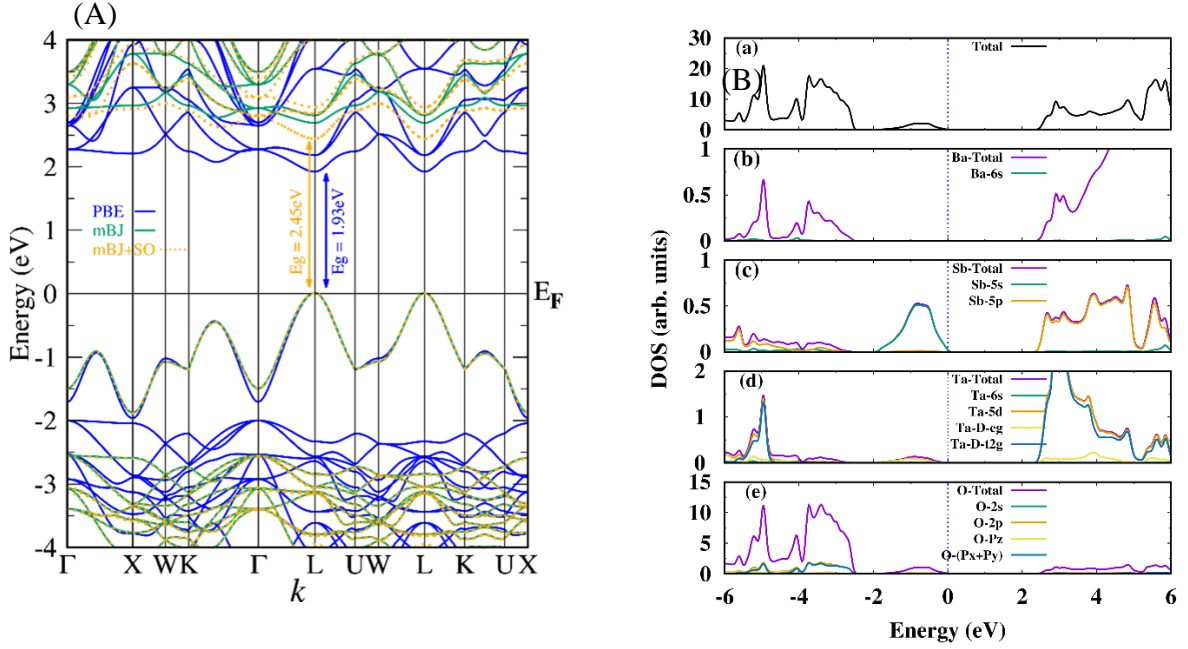

Fig. 3 Same as Fig. 2 except the compound which is $Ba_2SbTaO_6$.

From Fig. 2(B) and Fig. 3(B), it is observed that the energy bands within the energy range –6 eV to –2.6 eV are mainly formed by the 6s-electrons of Ba atoms with the small contributions of the 5p- and 2sp- electrons of Sb and O atoms, respectively. The valence band, which is just below the Fermi level ($E_F$), shows the 5s and 2s electrons of Sb and O atom, respectively. The bunches of the energy bands in the lower conduction region are resulted because of the 5s (6s) and 5d (4d) electrons of Nb (Ta) atoms, 5p electrons of Sb atoms, 6s electrons of Ba atoms, and 2s electrons of O-atoms. The bands within the energy range 2.6 eV to 6 eV are mainly constructed by the 4d orbitals of Nb (Ta), 6s orbitals of Ba, 5p-orbitals of Sb atoms with the small contributions of 5p-orbitals of Nb (Ta) and 5s-orbitals of Sb atoms.

### 3.4 Optical properties

The dielectric response of any material can be investigated using the dielectric function of the material. It is worth mentioning that the dielectric function of any material can be written as [50],

$$\varepsilon(\omega) = \varepsilon_1(\omega) + i\varepsilon_2(\omega). \tag{8}$$

Here $\varepsilon_1(\omega)$ and $2(\omega)$ correspond to the real and imaginary parts of the dielectric function, respectively. The variation of real and imaginary components of a dielectric tensor with the photon energy (up to 8 eV), for both compounds are plotted in Fig. 4(a) and Fig. (b), respectively. The real part of the dielectric function for the $Ba_2SbNbO_6$ ($Ba_2SbTaO_6$) compounds (Fig. 4(a)) increases gradually up to the 3.1 (3.2) eV and then it decreases up to the energy 3.8 (4.0) eV. Further, an increase in $\varepsilon_1(\omega)$ function is observed up to 4.8 eV and 5.2 eV for the examined compounds $Ba_2SbNbO_6$ and $Ba_2SbTaO_6$, respectively. The $\varepsilon_1(\omega)$ become zero at 6.2 (6.8) eV for $Ba_2SbNbO_6$ ($Ba_2SbTaO_6$) compound. The value of static constant $\varepsilon_1(0)$ for $Ba_2SbNbO_6$ and $Ba_2SbTaO_6$ is 4.6 and 4.0, respectively, which are per the Penn model [51]. The $\varepsilon_2(\omega)$ curve starts from 1.8 eV and 2.2 eV for $Ba_2SbNbO_6$ and $Ba_2SbTaO_6$, respectively (Fig. 4(b)). These values are closer to the band gap and are known as fundamental absorption edge. The fundamental absorption edges correspond to the optical transitions between VBM and [50]. Optical response of high temperature superconductors by full potential LAPW band structure calculations [50, 52]. In addition, various peaks are observed in $\varepsilon_2(\omega)$ spectra (Fig. 4b). These peaks are originated from the different inter-band transitions between the valence band and conduction band.

The variation in optical conductivity for both the studied compounds is shown in Fig. 4(c). The optical conductivity starts from the energy approximately at 3 eV for both the compounds and reaches up to the values $0.56 \times 10^4$ ($\Omega$ cm)$^{-1}$ and $0.6 \times 10^4$ ($\Omega$ cm)$^{-1}$ at energy 5.4 eV and 6.2 eV for $Ba_2SbNbO_6$ and $Ba_2SbTaO_6$, respectively. The non-vanishing optical conductivity values in the visible region show that the studied compounds can be used in various optoelectronic applications. Another important optical parameter, which illustrates the amount of light being absorbed by the material, is the absorption coefficient, $\alpha(\omega)$, of the material. The energy dependent absorption spectra for both compounds are presented in Fig. 4(d). From Fig. 4(d), absorption edges for $Ba_2SbNbO_6$ and $Ba_2SbTaO_6$ are observed at 1.8 and 2.4 eV, respectively. The significant intensity of $\alpha(\omega)$ is observed for the energy values greater than 3 eV which shows that both the studied compounds have potential to absorb the visible as well as UV radiations. The knowledge of the refractive index ($\eta$) and extinction coefficient helps to understand the optoelectronic properties of material which is essentially required for the practical applications of the material. The variation of refractive index ($\eta$) with photon energy for both the studied compounds are displayed in Fig. 5(a). The refractive index for $Ba_2SbNbO_6$ and $Ba_2SbTaO_6$ compounds, at 0 eV energy, are found to be 2.0 and 2.2, respectively, whereas the maximum value of ($\eta$) is observed as 2.8 (2.6) at 3.0 (6.2) eV energy for $Ba_2SbNbO_6$ ($Ba_2SbTaO_6$).

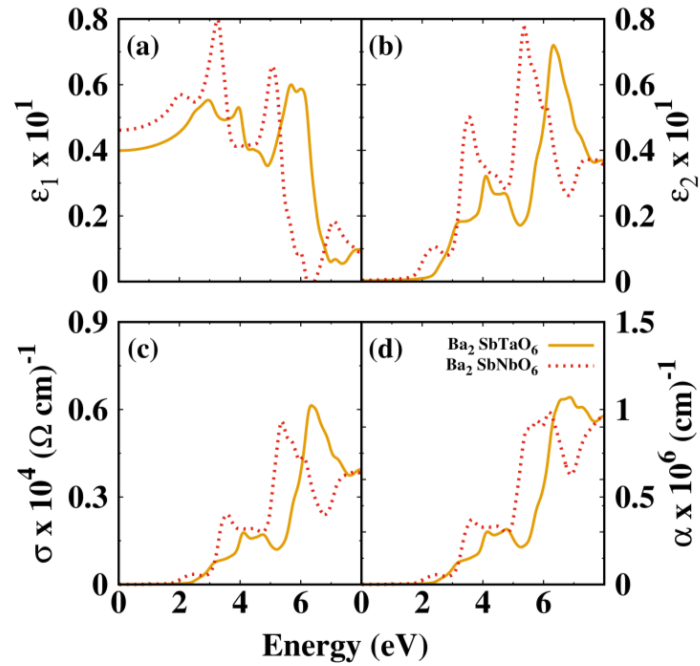

Fig. 4 Energy dependent (a) optical conductivity $\sigma(\omega)$, (b) absorption spectra $\alpha(\omega)$, (c, d) real and imaginary component of dielectric tensor for $Ba_2SbNbO_6$ and $Ba_2SbTaO_6$ compounds obtained from mBJ+SO calculations.

The energy dependent extinction coefficient, $k(\omega)$ which represents the damping of oscillation amplitude of incident electric field, is shown in Fig. 5(b) for the studied compounds. The reflectivity $R(\omega)$ of the studied compounds $Ba_2SbNbO_6$ and $Ba_2SbTaO_6$ have also been considered within the energy range 0–8 eV and are plotted in Fig. 5 (c). At 0 eV energy the reflectivity of $Ba_2SbNbO_6$ and $Ba_2SbTaO_6$ are observed as 11.5% and 14%, respectively, which confirms the semiconducting nature of the studied compounds. The reflectivity coefficient increases with increasing energy of the photons and attain a maximum value at 4.6 eV and 6.2 eV for $Ba_2SbNbO_6$ and $Ba_2SbTaO_6$, respectively.

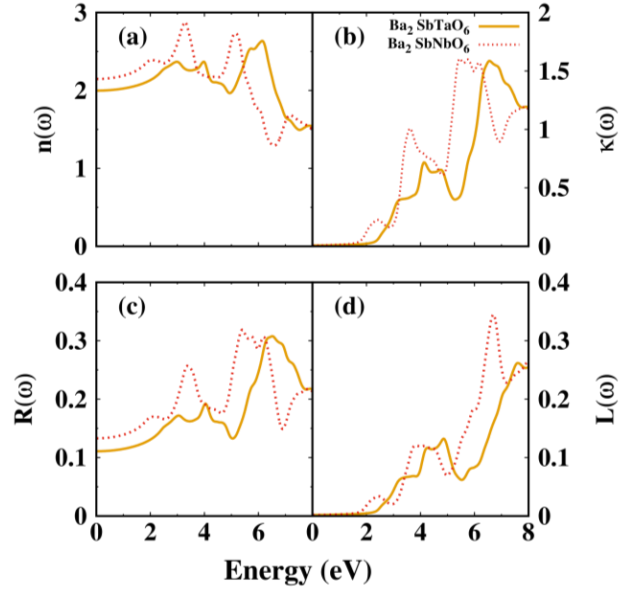

Fig. 5 Frequency dependent (a) refractive index $\eta(\omega)$, (b) extinction coefficient $\kappa(\omega)$, (c) reflectivity $R(\omega)$, and (d) energy loss function $L(\omega)$ for $Ba_2SbNbO_6$ and $Ba_2SbTaO_6$ compounds.

Electron energy loss function $L(\omega)$ is also a significant optical constant to understand the optical properties of material. It gives the information regarding the plasma frequency for the materials and the scattering of electrons passing through the materials. The variation of loss function with the energy is shown in Fig. 5 (d). The peaks in the graph of $L(\omega)$ represent the resonance condition and give the plasma frequency. At the energy above and below to plasma frequency, material shows dielectric and metallic/semiconducting nature, respectively. The highest peak for $Ba_2SbNbO_6$ and $Ba_2SbTaO_6$ are observed at 6.4 eV and 7.8 eV, respectively. The peaks available in the electron energy loss function correspond to the transition point for semiconducting to dielectric nature of $Ba_2SbNbO_6$ and $Ba_2SbTaO_6$.

## 3.5 Thermoelectric Properties

To check the thermoelectric performance of any substance, the figure of merit (ZT), the key parameter can be calculated using the Seebeck coefficient, electrical conductivity, and thermal conductivity. We have calculated these parameters of the compounds of $Ba_2SbNbO_6$ and $Ba_2SbTaO_6$ as a function of chemical potential $(\mu - \mu_F)$ eV over a wide range of temperature, 300-800 K. Fig. 6(a) and 7(a) depict the Seebeck coefficient for $Ba_2SbNbO_6$ and $Ba_2SbTaO_6$, respectively.

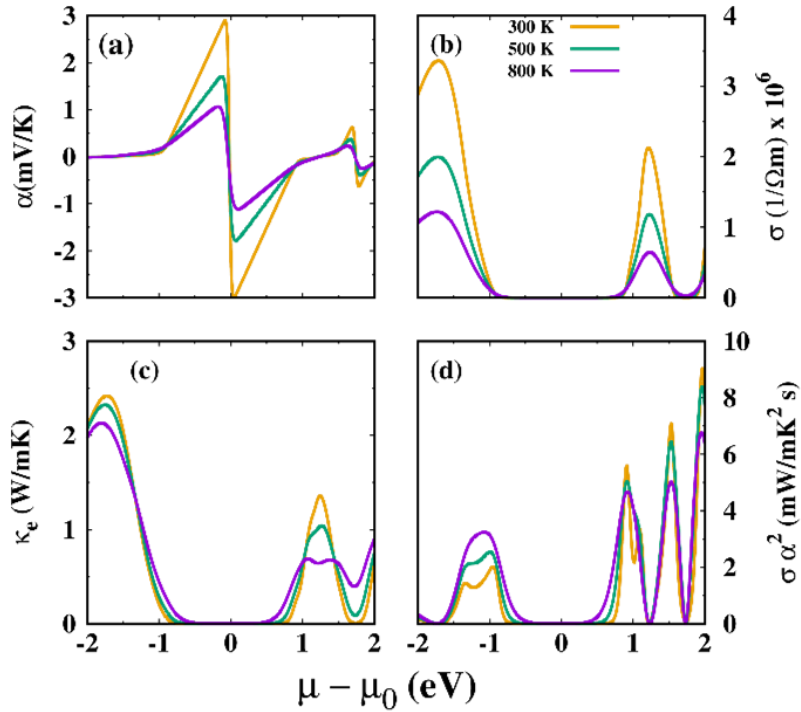

Fig. 6 Thermoelectric properties (a) Seebeck coefficient (b) Electrical conductivity (c) Electronic thermal conductivity and (d) Power factor for $Ba_2SbNbO_6$ as a function of chemical potential.

To characterize the types of dominant carriers in any system, the Seebeck coefficient modules are referred. We see from Fig. 6(a) that the Seebeck coefficient for electrons (holes) for $Ba_2SbNbO_6$ at 300K, 500K, and 800K is 298 µV/K (288 µV/K), 170 µV/K (178 µV/K), and 106 µV/K (111μV/K), respectively. Similarly, obtained values for Seebeck coefficient for $Ba_2SbTaO_6$ (Fig 7a) for the electron (holes) is 312 μV/K (301 μV/K), 241 μV/K (237 μV/K), and 150 μV/K (148 μV/K), at 300K, 500K, and 800K temperature values, respectively.

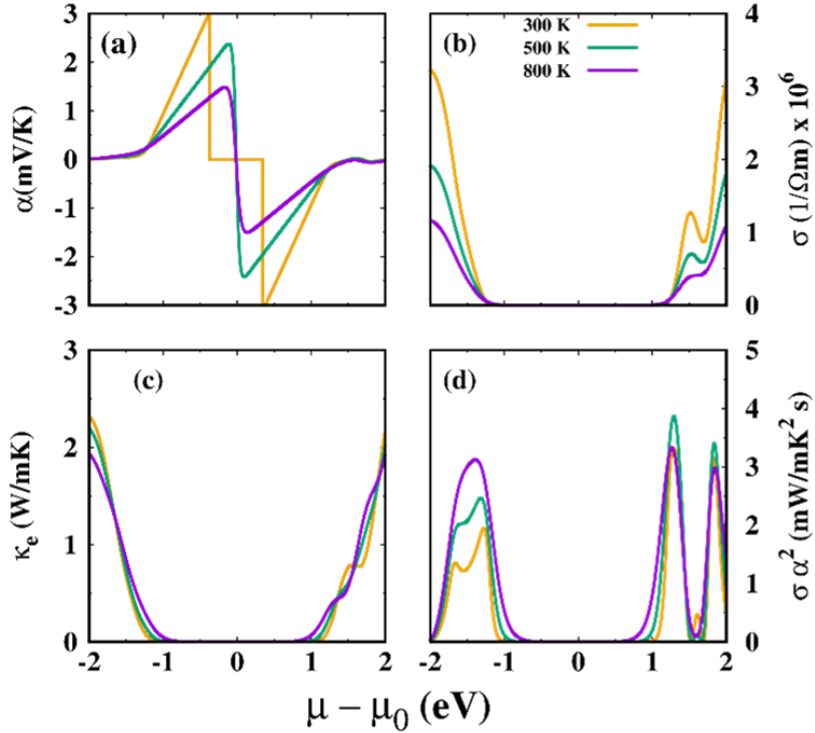

Fig. 7 Same as Fig. 6 except the compound which is $Ba_2SbTaO_6$.

We see from the values of the Seebeck coefficient that n-type carriers are dominated for both the compounds. Further, from Fig. 6(a) and 7(a), it is evident that $Ba_2SbTaO_6$ shows comparatively higher Seebeck coefficient values than $Ba_2SbNbO_6$, which is due to its wide bandgap than $Ba_2SbNbO_6$. We also notice that the Seebeck coefficient decreases with increasing the temperature value, which can be asserted to the excitation of charge carriers across VBM and CBM. The calculated values of effective mass (m*), using perturbation theory [53], for holes (electrons) for $Ba_2SbNbO_6$ and $Ba_2SbTaO_6$ is $0.36m_o$ $(0.98)m_o$ and $0.39m_o$ $(0.65)m_o$. We see $m^*_e$

> $m^*_h$ is also responsible for the higher Seebeck coefficient for electrons. BoltzTraP code, which is used in the present work for calculating thermoelectric properties, provides the electrical conductivity and electronic lattice thermal conductivity at a constant scattering relaxation time. To deduce the τ at various temperature values, we assumed a typical value for the τ (1x $10^{-14}$ s) at room temperature, 300 K, and then used in the relation $τ_T = 300 τ_{300}/T$ [54].

Electrical conductivity for $Ba_2SbNbO_6$ and $Ba_2SbTaO_6$ is shown in Fig. 6(b) and Fig. 7(b), respectively. The maximum value of electrical conductivity for n-type (p-type) chare carriers are obtained as $2.12 × 10^6$ ($3.36 × 10^6$) at chemical potential 1.21 (–1.70) ($μ − μ_F$) eV at 300 K temperature which decreases to $0.64 × 10^6$ ($1.21 × 10^6$) at chemical potential 1.23 (–1.73) ($μ − μ_F$) eV at 800 K for $Ba_2SbNbaO_6$.(Fig. 6b) while for $Ba_2SbTaO_6$ the maximum value of electrical conductivity for n-type (p-type) chare carriers are obtained as $3.09 × 10^6$ ($3.23 × 10^6$) at chemical potential 1.99 (–1.99) ($μ − μ_F$) eV at a 300 K temperature which decreases to $1.09 × 10^6$ ($1.16 × 10^6$) at chemical potential 1.99 (–1.99) ($μ − μ_F$) eV at 800 K (Fig. 7b). Larger values of electrical conductivity for holes than that of electrons is reflected from the behavior of the effective mass ($m^*_e > m^*_h$) as electrical conductivity is inversely proportional to the carriers' effective masses. We see that the electrical conductivity shows a decreasing trend with respect to temperature. This implies that at the higher temperature, the intrinsic phonon scattering mechanism is dominated. The electronic thermal conductivity is elucidated in Fig. 6(c) and Fig. 7(c) for $Ba_2SbNbO_6$ and $Ba_2SbNbO_6$, respectively. The electronic thermal conductivity of holes is also higher than that of electrons and it decreases with increasing temperature. This results from the coupled relationship between electrical conductivity and electronic thermal conductivity given by Wiedemann-Franz law ($k_e/σ$ is proportional with the temperature). Electrical conductivity combined with the Seebeck coefficient gives the power factor (PF) ($σS^2$). The power factor versus the chemical potential for Ba$_2$SbNbO$_6$ and Ba$_2$SbTaO$_6$ is shown in Fig. 6 (d) and 7 (d), respectively. Seebeck coefficient is a dominating quantity for PF. Higher Seebeck coefficient will result in higher PF. We see from Fig. 6 & 7(d) that the PF is higher for n-type charge carriers, which can be understood from the higher value of the Seebeck coefficient for electrons. The first peak of PF is observed around ~1 ($\mu - \mu_F$) eV chemical potential for Ba$_2$SbNbO$_6$ while for Ba$_2$SbTbO$_6$, it is noted near ~1.3 ($\mu - \mu_F$) eV for both the charge carriers. These values of chemical potential are quite realistic and could be realized practically. The maximum PF values for Ba$_2$SbNbO$_6$ for electrons (holes) is $5.6 \times 10^{-3}$ ($2 \times 10^{-3}$) at 0.91 (–0.95) ($\mu - \mu_F$) eV, $5.04 \times 10^{-3}$ ($2.53 \times 10^{-3}$) at 0.91 (–0.99) ($\mu - \mu_F$) eV, and $4.67 \times 10^{-3}$ ($3.2 \times 10^{-3}$) at 0.92 (–1.07) ($\mu - \mu_F$) eV for the 300K, 500K and 800K temperature values, respectively. Ba$_2$SbTaO$_6$ gives maximum PF for electrons (holes) $3.31 \times 10^{-3}$ ($1.94 \times 10^{-3}$) at 1.29 (–1.27) ($\mu - \mu_F$) eV, $3.38 \times 10^{-3}$ ($2.46 \times 10^{-3}$) at 1.29 (–1.31) ($\mu - \mu_F$) eV, and $3.24 \times 10^{-3}$ ($3.12 \times 10^{-2}$) at 1.26 (–1.38) ($\mu - \mu_F$) eV for the 300K, 500K and 800K temperature values, respectively. The PF is more promising for Ba$_2$SbNbO$_6$ within the pragmatic range of chemical potential. Finally, the ZT is calculated using the following relation:

$$ZT = \frac{\sigma T S^2}{K} \qquad (9)$$

where $K$ is the sum of electrical ($k_e$) and lattice thermal conductivity ($k_l$) and T is the temperature. BoltzTraP code employed here to assess thermoelectric data at zero $k_l$, resulted with variable magnitudes of ZT at various temperature ranges. However, as proposed by Slack [55], non-metallic crystals with high thermal conductivity have '$k_l$' as inversely proportional to the Grüneisen parameter and temperature. Presently recorded values of Grüneisen parameter, for Ba$_2$SbNbO$_6$ (1.65) and Ba$_2$SbTaO$_6$ (1.63), are comparatively twofold larger than that of the reported for the well-known thermoelectric material CaZrSe$_3$ (0.8) [56] and even in the vicinity of the values reported for Bi$_2$Te$_3$ (1.5) and Sb$_2$Te$_3$ (1.71) [57].

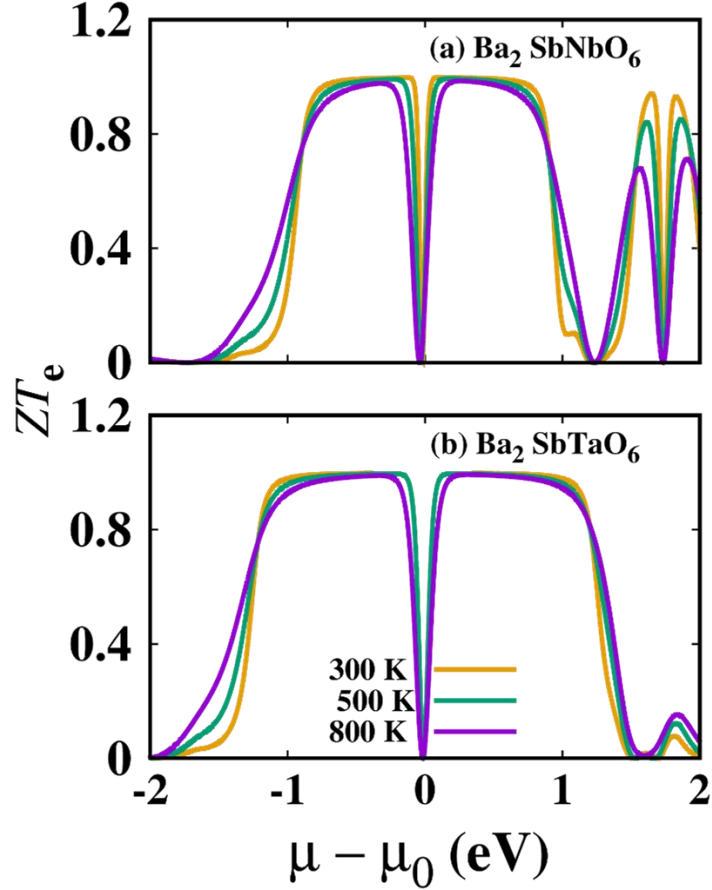

Fig. 8 Figure of merit (ZT) for Ba$_2$SbNbO$_6$ and Ba$_2$SbTaO$_6$ compounds.

Hence, with the moderate value of the Grüneisen parameter, the studied compounds are supposed to possess ultra-low lattice thermal conductivity as reported for the above-mentioned compounds. Thus, we are expecting that the ultra-low $k_l$ will not alter our reported ZT value significantly. The calculated values of ZT for electrons (holes) are 0.52 (0.29) at 0.88 (0.91) ) $(\mu - \mu_F)$ eV, 0.73 (0.59) at 0.83 (0.87) ) $(\mu - \mu_F)$ eV, and 0.85 (0.78) at 0.73 (0.78) ) $(\mu - \mu_F)$ eV for 300K , 500K and 800K, respectively for Ba$_2$SbNbO$_6$ while ZT for Ba$_2$SbTaO$_6$ are 0.38 (0.29) at 1.22 (1.23) )

($\mu - \mu_F$) eV, 0.67 (0.59) at 1.18 (1.19) ) ($\mu - \mu_F$) eV, and 0.83(0.78) at 1.09 (1.10) ) ($\mu - \mu_F$) eV for electrons (holes) at 300K, 500K and 800K, respectively (Fig. 8). From Fig. 8, it can be clearly understood that, ZT increases with an increase in temperature for both the studied compounds. Our calculated the melting point for $Ba_2SbNbO_6$ ($Ba_2SbTaO_6$) was 2041.25 K (2135.23K) which shows that the thermoelectric properties can be calculated at an even higher temperature range than reported here which could enhance the ZT.

### 3.6 SLME analysis

The photovoltaic (PV) performance of a material as an active absorber layer in a single junction solar cell can be evaluated by the "spectroscopic limited maximum efficiency (SLME)" introduced by Liping Yu and Alex Zunger [58]. SLME is a simulation model based on the improved Shockley-Queisser model that put forward the power conversion efficiency of a solar cell. SLME ($\eta$) is calculated as the ratio of the maximum output power density ($P_{max}$) to the total incident solar energy density ($P_{in}$), i.e. $\eta = P_{max}/P_{in}$. For SLME, the $P_{max}$ can be calculated using the current-voltage (*J-V*) characteristics of the solar cell as given by Eq. 10 [58-60],

$$P_{max} = \max(JV) = max\left\{\left(J_{SC} - J_0\left(e^{\frac{eV}{kT}} - 1\right)\right)\right\}, \qquad (10)$$

where *J*, $J_{SC}$, $J_0$, e, *V*, *k*, and *T* are the total current density, the short circuit current density, the reverse saturation current density, the elementary charge, the potential over the absorber layer, Boltzmann's constant, and the temperature of the device, respectively. The $J_{SC}$ and $J_0$ can be calculated using the absorptivity of the material [59]. The DFT calculated absorption spectra, global solar spectra (AM1.5G), fundamental ($E_g$) and direct allowed ($E_g^{da}$) band gap of the PV material are the main input parameters to analyse the SLME. Here we considered $E_g = E_g^{da}$.

In Fig, 9, we show the calculated SLME (%) for $Ba_2SbNbO_6$ (solid lines) and $Ba_2SbTaO_6$ (dashed lines) compounds at different temperatures and thickness values. We can see that the SLME is most affected by the device temperature, band gap, and thickness of the PV absorber material. SLME decreases from 26.8 % (12.9%) to 12.7 % (7.6 %) for $Ba_2SbNbO_6$ ($Ba_2SbTaO_6$) with increasing the device temperature from 293.15K to 800K, respectively while it gradually increases with the thickness of the material and the becomes constant after ~0.6µm.

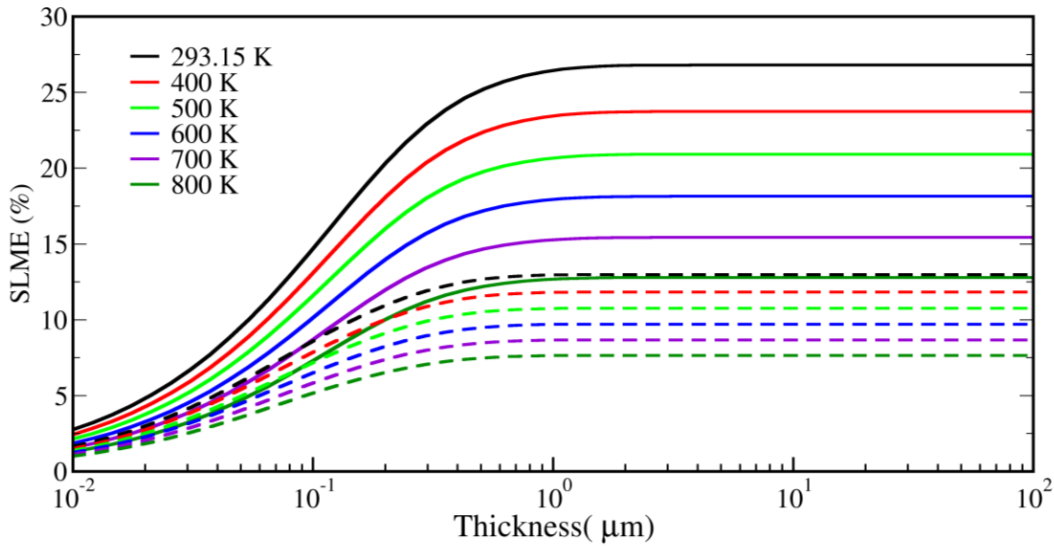

Fig. 9. Temperature and thickness dependent SLME for $Ba_2SbNbO_6$ (solid lines) and $Ba_2SbTaO_6$ (dashed lines) compounds.

The larger band gap of $Ba_2SbTaO_6$ (2.45 eV) comparative to the band gap of $Ba_2SbNbO_6$ (1.8 eV) causes the poor SLME of $Ba_2SbTaO_6$. Our calculated SLME analysis as shown in Fig. 9 screens out the $Ba_2SbNbO_6$ with a 26.8% power conversion efficiency (PCE) whereas the $Ba_2SbTaO_6$ is a very poor material with a PCE of 12.9% at 293.15 K which is the practically operating temperature for a solar cell. The 26.8% PEC of $Ba_2SbNbO_6$ is approaching the PEC of state-of-the-art inorganic PV cells (29.1%) and the Shockley-Queisser (S-Q) efficiency limit of ~33% [61-65] making $Ba_2SbNbO_6$ an appealing candidate for the single-junction solar cell.

**Conclusion**

In this paper, we have studied the structural, mechanical, electronic, and optical properties of lead-free double perovskite compounds $Ba_2SbXO_6$ (X = Nb and Ta) using density functional theory. Studied mechanical properties show the stability, ductile, and anisotropic nature of the studied compounds, revealing the suitability of these compounds in device fabrication. Computed electronic structure of $Ba_2SbXO_6$ (X = Nb and Ta) compounds shows the direct bandgap semiconducting nature of both the compounds with an energy gap of 1.80 eV and 2.45 eV for $Ba_2SbNbO_6$ and $Ba_2SbTaO_6$, respectively. Further from the computed absorption spectra, it is observed that both the studied compounds have good absorbing power for visible and UV radiation. Thermoelectric performance of the studied compounds has been assessed using power factor and figure of merit computed using the Seebeck coefficient, electrical conductivity, and the electronic thermal conductivity. The moderate values of the various calculated properties such as direct band gap, absorption spectra, power factor, and the figure of merit shows the potential of $Ba_2SbXO_6$ (X = Nb and Ta) compounds in optoelectronic, and thermoelectric applications. The PEC of 26.8% of $Ba_2SbNbO_6$ reveals its merits for photovoltaic devices.

**Acknowledgments:** Present work is financially supported by DST-SERB, New Delhi (India) under EMR schemes vide grant number EMR/2017/005534. A fruitful scientific discussion with Dr. R. Khenata, LPQ3M, University of Mascara, Mascara, Algeria is also acknowledged.